# A Novel Method for ECG Signal Classification via One-Dimensional Convolutional Neural Network

Xuan Hua[1,2], Jungang Han[2,3], Chen Zhao[4], Haipeng Tang[5], Zhuo He[4], Jinshan Tang[4], Qing-Hui Chen[6], Shaojie Tang7, Weihua Zhou[4]

[1]College of Electronic Engineering, Xi'an University of Posts and Telecommunications, Xi'an 710121 China

[2]Key Laboratory of Network Data Analysis and Intelligent Processing in Shaanxi, Xi'an 710121 China

[3]College of Computer Science and Technology, Xi'an University of Posts and Telecommunications, Xi'an 710121 China

[4]College of Computing, Michigan Technological University, Houghton MI 49931 USA

[5]School of Computing, University of Southern Mississippi, Long Beach 39560 USA

[6]Deprtment of Kinesiology and Integrative Physiology, Michigan Technological University, Houghton MI 49931 USA

[7]School of Automation, Xi'an University of Posts and Telecommunications, Xi'an 710121 China

Corresponding authors: Jungang Han (e-mail: hjg@xupt.edu.cn ) and Weihua Zhou (email: whzhou@mtu.edu )

**ABSTRACT** This paper presents an end-to-end ECG signal classification method based on a novel segmentation strategy via 1D Convolutional Neural Networks (CNN) to aid the classification of ECG signals. The ECG segmentation strategy named R-R-R strategy (i.e., retaining ECG data between the R peaks just before and after the current R peak) for segmenting the original ECG data into segments in order to train and test the 1D CNN models. The novel strategy mimics physicians in scanning ECG to a greater extent, and maximizes the inherent information of ECG segments. The performance of the classification models for 5-class and 6-class are verified with ECG signals from 48 records of the MIT-BIH arrhythmia database. As the heartbeat types are divided into 5 classes (i.e., normal beat, left bundle branch block beat, right bundle branch block beat, ventricular ectopic beat, and paced beat) in the MIT-BIH, the best classification accuracy, the area under the curve (AUC), the sensitivity and the F1-score reach 99.24%, 0.9994, 0.99 and 0.99, respectively. As the heartbeat types are divided into 6 classes (i.e., normal beat, left bundle branch block beat, right bundle branch block beat, ventricular ectopic beat, paced beat and other beats) in the MIT-BIH, the beat classification accuracy, the AUC, the sensitivity, and the F1-score reach 97.02%, 0.9966, 0.97, and 0.97, respectively. Meanwhile, according to the recommended practice from the Association for Advancement of Medical Instrumentation (AAMI), the heartbeat types are divided into 5 classes (i.e., normal beat, supraventricular ectopic beats, ventricular ectopic beats, fusion beats, and unclassifiable beats), the beat classification accuracy, the sensitivity, and the F1-score reach 97.45%, 0.97, and 0.97, respectively. The experimental results show that the proposed method achieves better performance than the state-of-the-art methods.

**KEYWORDS:** Electrocardiogram; convolutional neural networks; ECG signal classification; ECG segmentation strategy; support vector machine

## 1. Introduction

In recent years, the number of patients with cardiac diseases has significantly increased due to unhealthy eating habits and the lack of physical exercises [1]. And one of the most severe cardiac events is arrhythmia. However, it is hard to accurately diagnose arrhythmia with the existing hospital equipment. Hence, the best way to inspect arrhythmia is to measure relevant signals non-invasively[2] with an electrocardiogram (ECG) device. This method makes it easy to record a time series [3] of cardiac excitatory activities.

Before the prevalence of deep learning (DL), ECG signal classification mainly relied on the traditional algorithms that depend on feature extraction and classification by neural networks [4] [5] [6], support vector machine (SVM) [7] ,and hidden Markov model [8] [9] [10]. Those traditional algorithms mainly use a state transition matrix and the confusion probability matrix to predict and classify ECG signals, and show limited performance in diagnosing cardiac disease.

With the rapid development of artificial intelligence techniques, DL [11] [12] has been applied to ECG signal classification and achieved a good classification performance in recent years. In 2015, Meng et al. [13] proposed an ECG signal classification model, which was based on deep belief extraction features and Gaussian kernel nonlinear SVM. Their model achieved an accuracy of 98.49% by using the records of 46 patients in the MIT-BIH arrhythmia database (abbreviated as MIT-BIH hereafter) across 6 classes (i.e., normal beat, left bundle branch block beat, right bundle branch block beat, atrial premature beat, premature ventricular contraction, and paced beat). In their work, a segmentation strategy (i.e., retaining the heartbeat cycles just before or after the current heartbeat cycle) was used for ECG signal classification. In 2016, Zubair et al. [2] proposed a Convolutional Neural Network (CNN) based ECG beat classification system, which used the recordings of 44 patients in the MIT-BIH across 5 classes (i.e., normal beat, supraventricular ectopic beat, ventricular ectopic beat, fusion beat, and unknown beat) and achieved an accuracy of 92.7%. In their work, in order to learn the ECG beat patterns, a fixed number of samples (e.g., 100 samples) on both sides of the current R peak were extracted. In 2016, Kiranyaz et al. [3] proposed a real-time patient-specific ECG signal classification based on an adaptive 1-D CNN. Experiments were performed on the recordings of 44 patients in the MIT-BIH. Their algorithm achieved a superior classification performance for the detection of ventricular ectopic beats and supraventricular ectopic beats. However, they did not illustrate their evaluation criteria. In their work, in order to learn the morphological structure of the beat signals, equal number of samples from both sides of the current R peak were also used for ECG signal classification. In 2016, Al Rahhal et al. [12] proposed a DL approach of active classification for ECG signals from the records of 44 patients in the MIT-BIH. They learned deep features in an unsupervised way by using sparse de-drying self-encoding. As the feature learning was completed, a softmax layer was added at the top of the hidden layer to form a so-called deep neural network (DNN). In their work, all ECG signals were first preprocessed using a 200ms width median filter to remove P wave and QRS complex, then a 600ms width median filter to remove T wave. They extracted the ECG waveform and features by the ecgpuwave software, and a fixed length was used as the ECG segmentation strategy. Also, they did not give their evaluation criteria and only stated that their method provided a significant accuracy improvement. In 2017, Kan et al. [13] proposed a novel model incorporating automatic feature abstraction and a DNN classifier (with an encoder layer of the stacked denoising auto-encoder (SDA) and a softmax layer) for ECG signal classification. The heartbeat data samples were divided into 4 classes (i.e., normal beat, supraventricular ectopic beat, ventricular ectopic beat, and fusion beat), and the best accuracy was about 97.5% for the recordings of 44 patients in the MIT-BIH. In their work, 700ms data window, centered at the current R peak (i.e., 300ms before and 400ms after) was used to segment each heartbeat. Obviously, the segmentation strategy does not guarantee a complete heartbeat cycle, and may lose some information in a heartbeat cycle, leading to an unsatisfactory ECG classification performance. In 2018, Al Rahhal et al. [14] proposed a method which was tested on the MIT-BIH arrhythmia, the INCART and the SVDB databases and achieved better results in the detection of ventricular ectopic beats (VEB) and supraventricular ectopic beats (SVEB), which are comparable to the state-of-the-art methods. However, this method could only detect two types of beats and is not comparable to our proposed method below.

To improve the performance of ECG signal classification, we propose an ECG segmentation strategy (i.e., the R-R-R segmentation) for ECG signal classification with an end-to-end 1D CNN model. We use the R-R-R strategy for cutting the original ECG data into segments that carry the diagnosing information for training the models. Especially, this strategy is suggested by an experienced expert to mimics the experiences that physicians used in examining ECG to a greater extent and improves the classification accuracy with minor extra computation. Table 1 shows the existing strategies for ECG segmentation.

Table 1. The existing strategies for ECG segmentation

| Year | ECG segmentation strategy |
|---|---|
| **This paper** | **The R-R-R segmentation strategy** |
| 2018[14] | All ECG signals segmented with the same length |
| 2017[13] | Centered at the current R peaks (300ms before and 400ms after) |
| 2016[2] | Fixed number of 100 samples on both sides from the current R peak |
| 2016[3] | Equal number of samples from both sides of the current R peak |
| 2015[15] | Retaining the beat cycles just before and after the current beat cycle |
| 2012[16] | Centered at the current R peaks (300ms before and 400ms after) |

## 2. Materials and Methods

### A. Database

The ECG signals from the 48 recordings of 47 patients in the MIT-BIH dataset [17] [18] [19] were used for all experiments in this work. A band pass filter at 0.1–100Hz was applied to each ECG signal [20], and all sampling frequencies were unified to 360 Hz [21]. Each record contained 30 minutes of data segmented from 24 hours of data acquired with two leads. And the two leads were the modified limb lead II and one of the modified leads V1, V2, V4, or V5 [20]. For cardiologists, any abnormality in heart rate or changes in recorded ECG morphological patterns can be detected as a marker for arrhythmias [3]. This database was fully annotated by cardiologists, wherein the annotation information included the locations of the R peaks and the types of cardiac events. The heartbeat types in the MIT-BIH are listed in Table 2.

Table 2. The cardiac event types with corresponding codes in the MIT-BIH

| Code | Label_store | Symbol | Description | Number of beats | Total |
|---|---|---|---|---|---|
| **1** | **1** | **N** | **Normal beat** | **75016** | **75016** |
| **2** | **2** | **L** | **Left bundle branch block beat** | **8072** | **8072** |
| **3** | **3** | **R** | **Right bundle branch block beat** | **7256** | **7256** |
| **5** | **5** | **V** | **Premature ventricular contraction** | **7130** | **7130** |
| **12** | **12** | **/** | **Paced beat** | **7024** | **7024** |
| 0 | 4 | a | Aberrated atrial premature beat | 150 | 4948 |
| | 6 | F | Fusion of ventricular and normal beat | 803 | |
| | 7 | J | Nodal (junctional) premature beat | 83 | |
| | 8 | A | Atrial premature contraction | 2544 | |
| | 9 | S | Premature or ectopic supraventricular beat | 2 | |
| | 10 | E | Ventricular escape beat | 106 | |
| | 11 | j | Nodal (junctional) escape beat | 229 | |
| | 13 | Q | Unclassifiable beat | 33 | |
| | 34 | e | Atrial escape beats | 16 | |
| | 38 | f | Fusion of paced and normal beat | 982 | |

### B. ECG Segmentation

ECG segmentation and location detection are key for ECG signal classification. Howver, the detection of the locations of R-peaks [22] [23] is beyond the scope of this paper. We directly used the data that already had R-peaks location indexed.

According to the physician's recommendation and referencing to the previous ECG segmentation strategies (as shown in Table 1), we used the R-R-R strategy to cut the raw ECG data into segments and only retained the R-peaks before and after the current R peak. On the one hand, our method mimiced physicians in scanning ECG to a greater extent; on the other hand, comparing with the previous ECG segmentation strategies, each segment acquired by our method always contained more signals than a complete heartbeat cycle. This segmentation strategy facilitates the network model training to abstract the latent features of ECG signals in the MIT-BIH with minor extra computation.

### C. ECG Classification

Our study selected all 48 recordings of all 47 patients in the MIT-BIH database, which included the class of paced beat that was removed by the previous research. In fact, the signal of the paced beat is very similar to the signal of the normal beat, making it more difficult to discriminate between them. We can attack this challenge by our novel segmentation strategy.

Moreover, according to the recommended practice from AAMI, record 102, 104, 107, and 217 are excluded because these beats do not have completely sufficient signal quality in diagnosing cardiac diseases [2]. AAMI recommends that each ECG beat be classified into the following five heart beat types: N (normal beats), S (supraventricular ectopic beats), V (ventricular ectopic beats), and F (fusion beats), and Q (unclassifiable beats). The ECG class description based on AAMI standards is given in Table 3. At the same time, following the clinician's recommendation, two classifiers were designed, i.e. the 5-class classifier and the 6-class classifier. The former divided the heartbeat events into 5 classes (i.e., normal beat, left bundle branch block beat, right bundle branch block beat, ventricular ectopic beat, and paced beat, simultaneously see Fig. 1). The latter divided the heartbeat events into 6 classes, which included the aforementioned 5 classes and an extra class 0 (i.e. the remaining 10 codes in MIT-BIH).

Table 3. ECG Class description using AAMI standard

| AAMI Classes | MIT-BIH heartbeat types | | | |
|---|---|---|---|---|
| **Normal beat (N)** | Normal beat (N) | Left bundle branch block beat (L) | Right bundle branch block beat (R) | Atrial escape beat (e) | Nodal (junctional) escape beat (j) |
| **Supraventricular ectopic beat (S)** | Atrial premature beat (A) | Aberrated atrial premature beat (a) | Nodal (junctional) premature beat (J) | Supraventricular premature beat (S) |
| **Ventricular ectopic beat (V)** | Premature ventricular contraction (V) | Ventricular escape beat (E) | | |
| **Fusion beat (F)** | Fusion of ventricular and normal beat (F) | | | |
| **Unknown beat (Q)** | Paced beat (/) | Fusion of paced and normal beat (f) | Unclassified beat (Q) | |

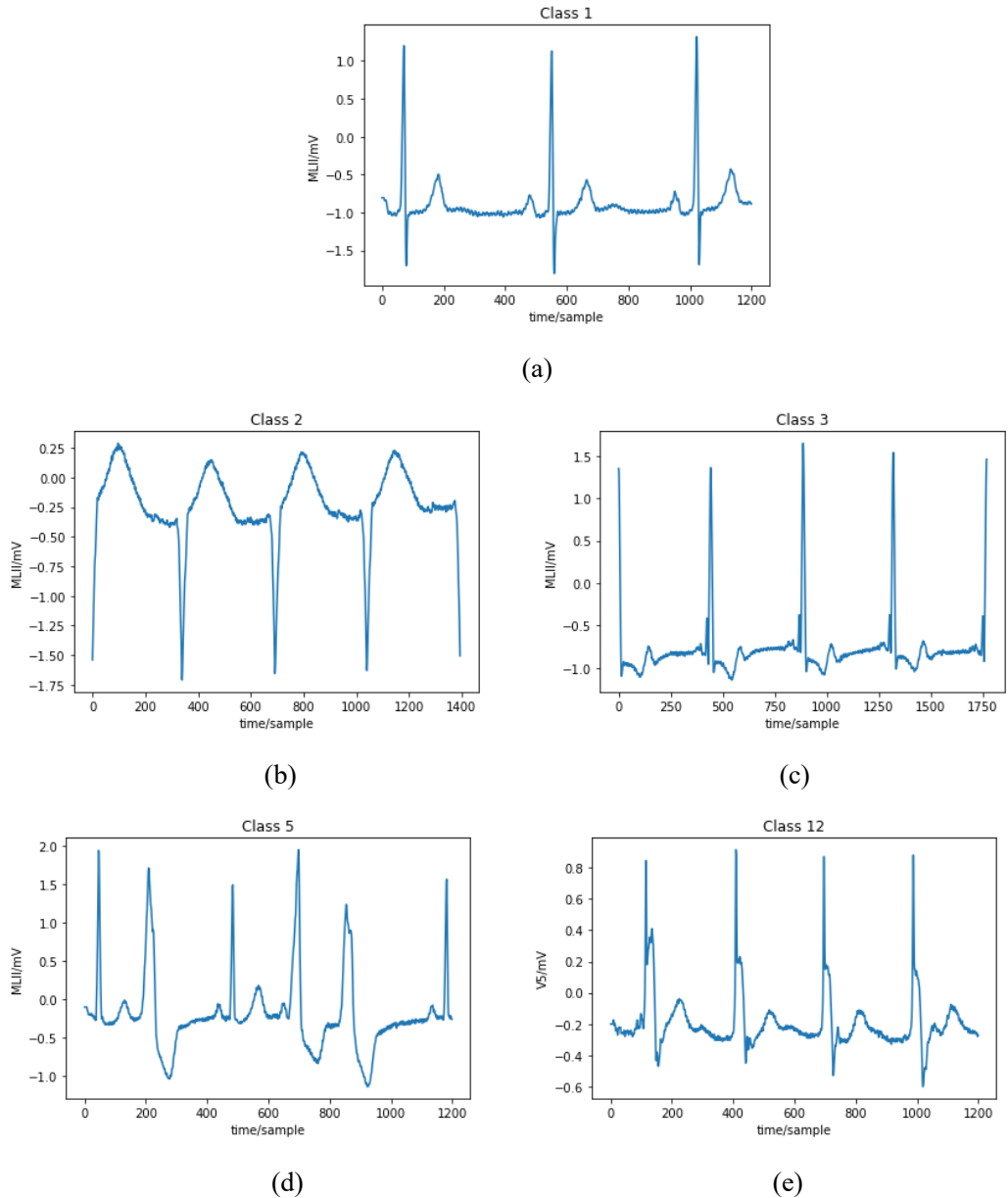

Figure 1. 5 classes for 5-class classifier (a) class 1 (normal beat), (b) class 2 (left bundle branch block beat), (c) class 3 (right bundle branch block beat), (d) class 5 (premature ventricular ectopic beat) and (e) class 12 (paced beat).

D. Data Preprocessing

The statistic results for all codes are shown in Fig. 2, where some cardiac event types are of inferior statistics in the MIT-BIH. Both Fig. 2 and Table 2 show that the statistics of samples with different heartbeat types are extremely unbalanced. If we train a model with the unbalanced ECG data directly, the performance would be extremely low. It is necessary to preprocess the ECG data by balancing and complementing them.

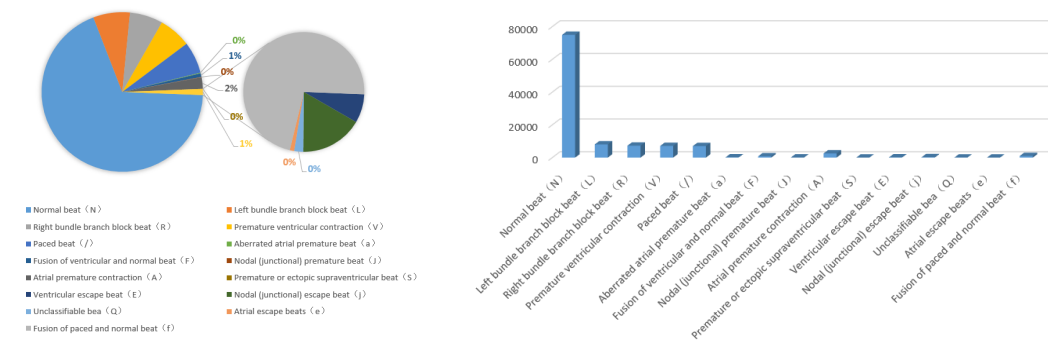

(a) (b)

Figure 2. The statistics for cardiac event types according to the annotated codes in MIT-BIH arrhythmia database, (a) the Pie Chart and (b) the Histogram

D1. Data Balance

From Table 2, it can be seen that the number of the ECG segments in the normal beat is extremely large, around 10 times the number of those in the left bundle branch block beat, the right bundle branch block beat, the premature ventricular ectopic beat or the paced beat. Normal heartbeats accounted for 71.79% (n = 75016) [24] of the total number of ECG segments. The hearts were divided into 5 classes(normal beat, left bundle branch block beat, right bundle branch block beat, ventricular ectopic beat, and paced beat) or 6 classes in the MIT-BIH. Using the sub-sampling rules proven by [25], we randomly sampled about 10% (n = 7475) ECG segments out of the normal beat, which was roughly equal to the number of those in the left bundle branch block beat, the right bundle branch block beat, the premature ventricular ectopic beat or the paced beat. The number of heartbeats in each sample set is listed in Table 4 and Table 5 in the MIT-BIH. The number of heartbeats in each sample set is listed in Table 6 by according to the AAMI recommendation. At the same time, we randomly sampled about 10% ECG segments out of the normal beat in order to balance other beats. We randomly selected 80% heartbeats from the total samples in the whole experiment as the training set, and the remainder of the samples was used as the testing set. The experiments show that model performance can be significantly improved by training with balanced data.

Table 4. The data partition in 5-class classification in MIT-BIH

| Symbol | N | L | R | V | / | Total |
|---|---|---|---|---|---|---|
| Code | 1 | 2 | 3 | 5 | 12 | |
| Training set | 5980 | 6456 | 5688 | 5748 | 5744 | 29616 |
| Testing set | 1495 | 1614 | 1422 | 1437 | 1436 | 7404 |

Table 5. The data partition in 6-class classification in MIT-BIH

| Symbol | N | L | R | V | / | Other | Total |
|---|---|---|---|---|---|---|---|
| Code | 1 | 2 | 3 | 5 | 12 | 0 | |
| Training set | 6000 | 6360 | 5788 | 5720 | 5536 | 6500 | 35904 |
| Testing set | 1500 | 1590 | 1447 | 1430 | 1384 | 1625 | 8976 |

Table 6. The data partition according to the AAMI recommendation

| Symbol | N | S | V | F | Q | Total |
|---|---|---|---|---|---|---|
| Code | 1 | 4 | 3 | 0 | 2 | |
| Training set | 7192 | 2168 | 5948 | 556 | 6500 | 22364 |
| Testing set | 1798 | 542 | 1487 | 139 | 1625 | 5591 |

D2. Data complementing

According to the matrix multiplication rule of a fully connected layer, the input layer needs a fixed input length. However, in our ECG segmentation method, the length of each R-R-R segment was not equal. To mend this, we chose a window length to be large enough (e.g., 2700 here) and placed the current R peak at the center of the window. If the length of the input ECG segment was less than the window's length, the input R-R-R segment must be complemented by zero padding at both sides. We noted that the maximum length of all R-R-R segments acquired by us from the MIT-BIH database was about 2600. So, all R-R-R segments can be added to the data window with a fixed length 2700, which was convenient for training and testing models.

E. Convolutional Neural Networks

CNN [26] is feed-forward and widely used for feature extraction. It does not require too much preprocessing for the original information, and can automatically generate high-level features by training [24]. To classify the signal with a short duration, 1D CNNs have become popular in various signal processing applications such as structural damage detection, high power engine fault monitoring, and real-time monitoring of high-power circuitry. Two recent studies have utilized 1D CNNs for damage detection and get an accuracy of 93.61% [27] [28].

The 1D CNNs are relatively easier to train and offer minimal computational complexity while achieving good performance. The ECG segments are essentially 1-D data, and we used the 1D-CNN network to acquire as many informative features as possible to better train the model.

CNN generally consists of one input layer, several convolution layers, several pooling layers, and a fully connected layer. The model we used contained three convolutional layers, each of which extracted different levels of features from the ECG segments. The convolutional layers were used to effectively extract multi-level features from ECG segments, and each convolutional layer could be considered as a fuzzy filter [24] that enhanced the characteristics of the original signal and reduced the noise. The weight sharing of convolutional kernel can effectively reduce the number of training parameters and model complexity. According to the principle of local correlation and retaining useful information effectively while reducing the dimensions of data, the pooling layers were added to our model to reduce the number of parameters in the fully connected layer and prevented the network from overfitting. After multiple convolutional and pooling layers, the fully connected layer integrated and normalized highly abstracted features. The normalized features were finally classified by the Softmax classifier in the output layer. As a consequence, both feature extraction and classification operations were fused into one process that can be optimized the classification performance.

Compared with the 1D-CNN network proposed in [2] and [3], the input data length of the first layer in our network far exceeded that in [2] and [3]. Therefore, each segment obtained by our method always contained more information than one whole heartbeat cycle, which can maximize the inherent information of ECG segments. In conventional 1D CNNs, the input layer is a passive layer that receives the raw 1D signal and the output layer is a Multilayer Perceptron layer with the number of neurons equal to the number of classes. Our CNN architecture was a 1D-CNN that included convolution kernels with different sizes, and the input data length was set to 2700, which included more information than that of previous works. The parameter settings are shown in Fig. 3. The kernel sizes used for convolutional layer 1, 2 and 3 were set to 5, 10 and 15, respectively, and the sub-sampling factor was 5. During CNN training, the model performance was evaluated at regular time intervals with a test set, and the model with the best test result was saved rather than waiting for the model to be fully trained.

During the training process, the features within the fully connected layer should satisfy the classification criteria. The convolutional layer and the pooling layer were adjusted and optimized to further satisfy the classification criteria. In addition, the rectifier linear unit (ReLU) [24] was used as the activation function and the mean square error was used as the loss function of the CNN model. The learning rate was initiated with 0.0001 and was automatically attenuated along with the learning process until the model converged. To refine the parameters in CNN, we chose Adam optimizer [29], which can automatically adjust the learning rate, and the weight update was not affected by the gradient scaling. The set of CNN parameters that lead to the minimum training error and the maximum testing recall is shown in Table 7.

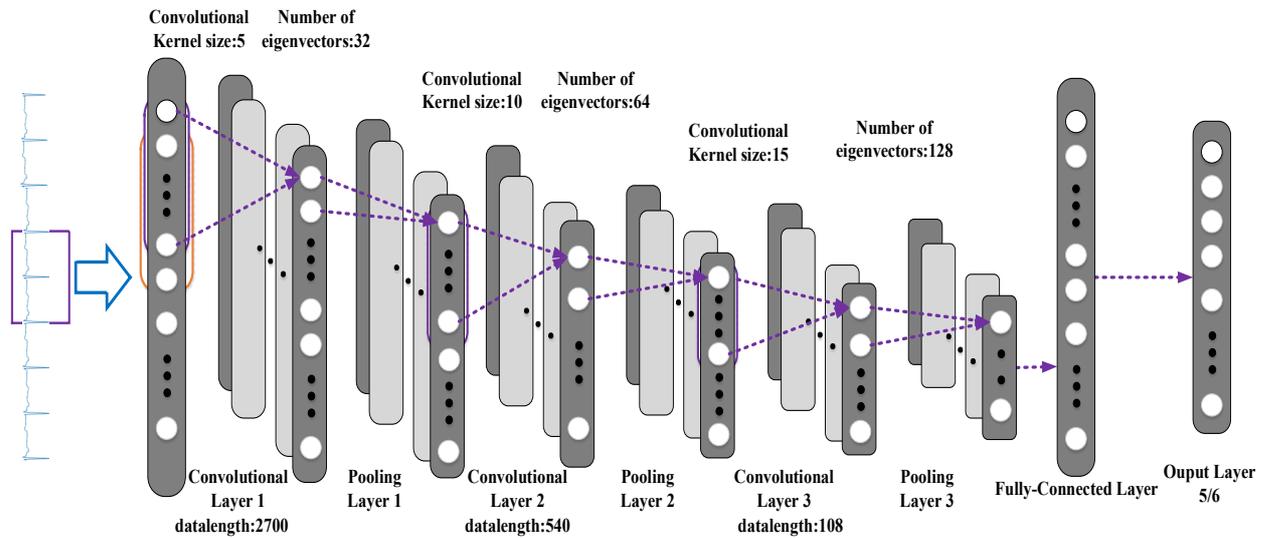

Figure 3. The architecture and parameters of the CNN used for ECG signal classification

Table 7. The setting of CNN model parameters designed for the ECG classification

| The architecture of CNN | Parameters Setting |
| --- | --- |
| Learning rate initial value | 0.0001 |
| The first convolutional layer kernel size | 5 |
| No. of feature maps in the first convolutional | 32 |
| The first sampling layer kernel size | 5 |
| The second convolutional layer kernel size | 10 |
| No. of feature maps in the second convolutional | 64 |
| The second sampling layer kernel size | 5 |
| The third convolutional layer kernel size | 15 |
| No. of feature maps in the third convolutional | 128 |
| The third sampling layer kernel size | 5 |
| No. of neurons in the fully connected layer | 2688 |
| Epoch number | 83 |

F．Evaluation metrics

Confusion matrix was generated by true positive (TP), false positive (FP), false negative (TN) and false negative (FN) [30] to describe the statistic relationship between the actual and predicted classes of ECG segments and evaluate the performance of a classifier. In addition, accuracy, sensitivity, F1-score, and AUC of the receiver operating characteristic (ROC) curve[31] were also used to evaluate ECG classification performance, as shown in the following 3 equations.

$$\text{Precision} = \frac{TP}{TP+FP}$$

$$\text{Sensitivity} = \frac{TP}{TP+FN}$$

$$F1 = \frac{2\,TP}{2\,TP+FP+FN}$$

False positive rate and true positive rate were used as the abscissa and ordinate of the

Cartesian coordinate system, respectively, to obtaion the ROC curve [31]. AUC was calculated from the ROC curve, and the classification performance [30] can be qualitatively judged according to the range of AUC, as shown in Table 8.

Table 8. The classification performance is qualitatively judged according to the range of AUC

| The range of AUC | Classification performance |
|---|---|
| 1.0 | Perfect |
| 0.90~1.00 | Excellent |
| 0.80~0.90 | Good |
| 0.70~0.80 | Medium |
| 0.60~0.70 | Poor |
| 0.50~0.60 | Failure |

3. Experimental Results

In this work, all experimental results with CNN and evaluation metrics were obtained using 7-fold cross-validations. The CNN was implemented in Keras on Linux running on a graphics processing unit (GPU) (GTX 1080 Titan XP). The tendency curves of cthe lassification errors and accuracies are shown in Figs 4 and 5 for the cases of 5 and 6 classes, respectively. The experimental results with CNN will be described later.

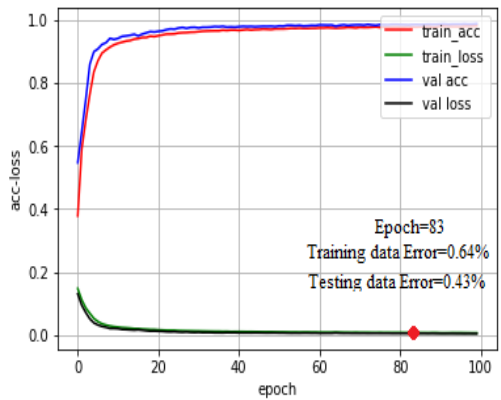
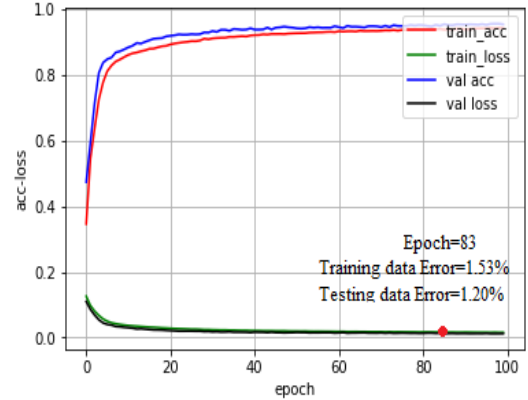

Figure 4. Classification errors with 5 classes    Figure 5. Classification errors with 6 classes

A1.The 5-class classifier in the MIT-BIH

Expereiments were performed for both 5-class and 6 class classifiers on the MIT-BIH database. For the 5-class classifier, the cardiac event types were divided into 5 classes. The confusion matrix of the model's performance for 5-class classifier is shown in Figs 6 and 7, the ROC curves of the model are shown in Figs 8 and 9, and the quantitative evaluation of the experimentsfor    5-class classifier is shown in Table 8.

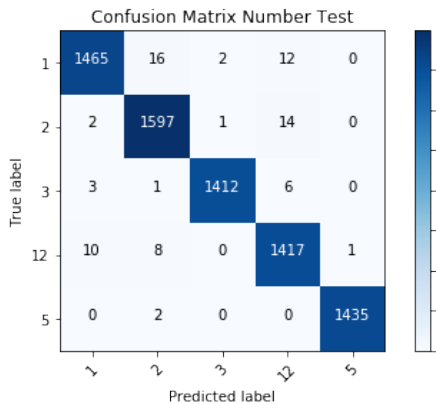
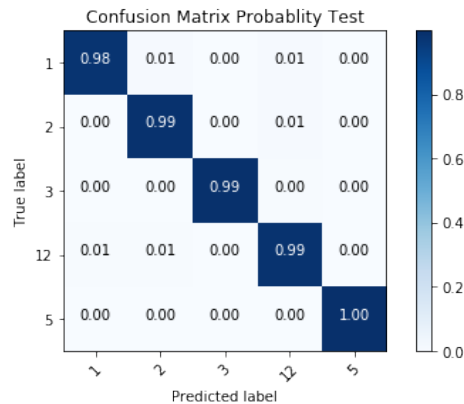

Figure 6. Confusion Matrix (Quantity)    Figure 7. Confusion Matrix (Probability)

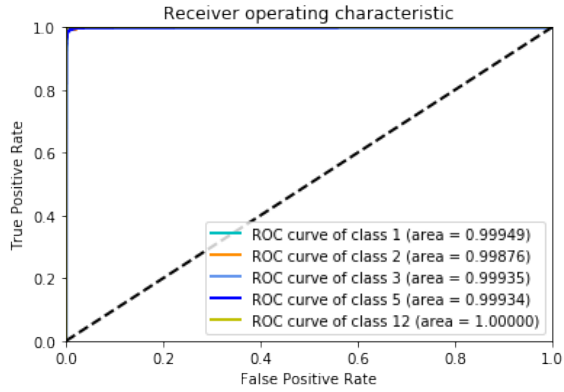
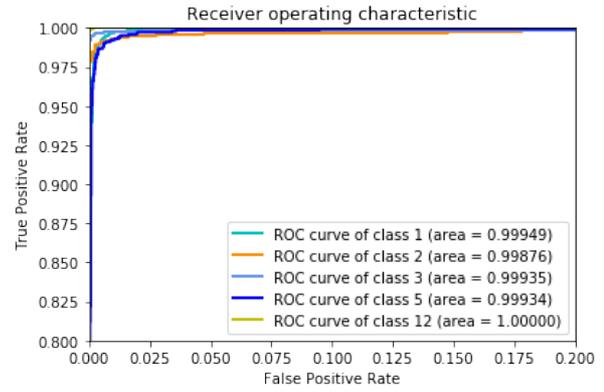

Figure 8. ROC curve (Global)　　　　　　　Figure 9. ROC curve (Local)

Table 8. The evaluation of the experiment with CNN as the cardiac event types are divided into 5 classes in MIT-BIH

| Code | Symbol | Accuracy | Sensitivity | F1-score |
|---|---|---|---|---|
| 1 | N | 0.99 | 0.98 | 0.98 |
| 2 | L | 0.98 | 0.99 | 0.99 |
| 3 | R | 1.00 | 0.99 | 1.00 |
| 12 | V | 0.98 | 0.99 | 0.98 |
| 5 | / | 1.00 | 1.00 | 1.00 |
| Average | | 0.99 | 0.99 | 0.99 |

A2. The 6-class classifier in the MIT-BIH

As the cardiac event types are divided into 6 classes, the confusion matrices for the experiment are shown in Figs 10 and 11, the ROC curves of the model are shown in Figs 12 and 13, and the quantitative evaluation of the experiment with CNN, is shown in Table 9.

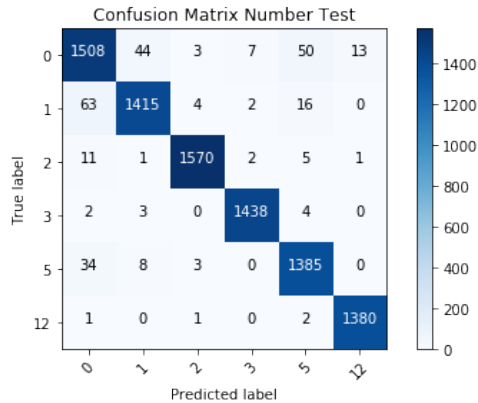
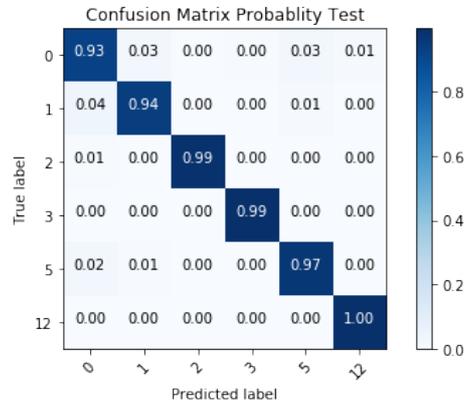

Figure 10. Confusion Matrix (Quantity)　　　Figure 11. Confusion Matrix (Probability)

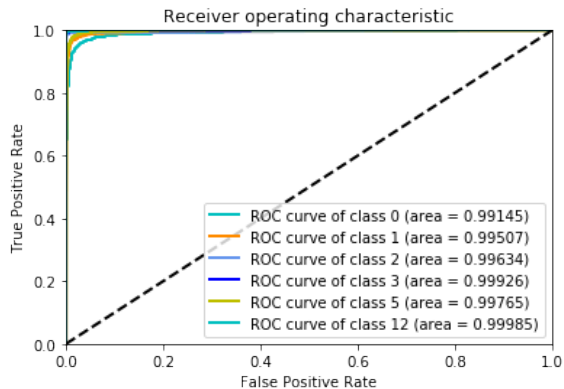
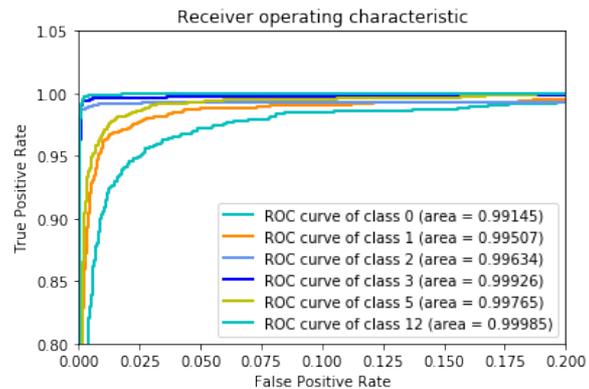

Figure 12. ROC curve (Global)　　　　　　Figure 13. ROC curve (Local)

Table 9. The evaluation of the experiment with CNN as the cardiac event types are divided into 6 classes in

MIT-BIH

| Code | Symbol | Accuracy | Sensitivity | F1-score |
|---|---|---|---|---|
| 0 | Other | 0.93 | 0.93 | 0.93 |
| 1 | N | 0.96 | 0.94 | 0.95 |
| 2 | L | 0.99 | 0.99 | 0.99 |
| 3 | R | 0.99 | 0.99 | 0.99 |
| 12 | V | 0.95 | 0.97 | 0.96 |
| 5 | | 0.99 | 1.00 | 0.99 |
| Average | | 0.97 | 0.97 | 0.97 |

A3. The 5-class classifier by the AAMI recommendation

As the cardiac event types are divided into 5 classes by the AAMI recommendation, the confusion matrices for the experiment are shown in Figs 14 and 15, and the quantitative evaluation of the experiment with CNN is shown in Table 10.

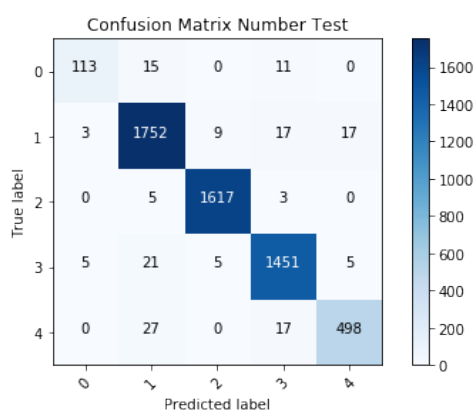
Figure 14. Confusion Matrix (Quantity)

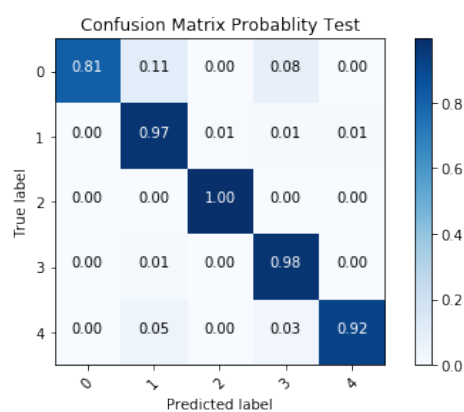
Figure 15. Confusion Matrix (Probability)

Table 10. The evaluation of the experiment with CNN as the cardiac event types are divided into 5 classes by the AAMI recommendation

| Code | Symbol | Accuracy | Sensitivity | F1-score |
|---|---|---|---|---|
| 1 | N | 0.96 | 0.97 | 0.97 |
| 4 | S | 0.96 | 0.92 | 0.94 |
| 3 | V | 0.97 | 0.98 | 0.97 |
| 0 | F | 0.93 | 0.81 | 0.87 |
| 2 | Q | 0.99 | 1.00 | 0.99 |
| Average | | 0.97 | 0.97 | 0.97 |

4. Discussions

We divided the types of cardiac events in the MIT-BIH into 5 and 6 classes, respectively. In the latter case, the other class contains all the remaining 10 cardiac event types except the normal beat, the left bundle branch block beat, the right bundle branch block beat, the ventricular ectopic beat, and the paced beat. The inter-class variation in other class is relatively huge, thus making it difficult to train the model. The performance of the 6-class classifier is lower than that of the 5-class classifier. The purpose of dividing the cardiac event types into 5 classes is to verify the possible best performance in an ideal situation, but this dividing strategy is difficult to use for an actual clinical circumstance. Dividing the cardiac event types into 6 classes is consistent with clinical practice, since arbitrary cardiac event type may happen.

Inspired by the way that physician diagnoses cardiac event types with the aids of ECG, we proposed R-R-R segmentation strategy for cutting the original ECG data into segments for training and testing models. There are two reasons for this new segmentation strategy: it mimics physicians in scanning ECG for diagnosing cardiac event types to a greater extent and each segment contains extra data beyond a complete beat cycle and improves the final ECG classification performance with minimal extra computation.

In recent years, researchers have continued to explore the classification of ECG signals. And the evaluation metrics in previous studies are focused on accuracy, as shown in Table 14. We use the accuracy, sensitivity, and F1-score performance measures for classification. Accuracy alone is not a good performance measure as we are working with biased data. This means that we have more benign ECG signal data than malign ECG signal data in the training set of data. In order to make a more accurate assessment of an experimental result, multiple evaluation metrics are needed, and F1-score is a synthetic metric, in our experiments. We focus on accuracy, sensitivity, and F1-score to evaluate the results of the experiment, as shown in Table 14. The table clearly show that the proposed method can achieve the best performance.

Table 14. The comparison of performance for different methods

| Year | Data Selection | Accuracy | Sensitivity | F1-score |
|---|---|---|---|---|
| **This paper** | **(5 classes) from 48 recordings in MIT-BIH** | **99.24%** | **0.99** | **0.99** |
| | **(6 classes) from 48 recordings in MIT-BIH** | **97.02%** | **0.97** | **0.97** |
| | **(5 classes) form 48 recordings by AAMI** | **97.45%** | **0.97** | **0.97** |
| 2018[14] | for VEB and SVEB in MIT-BIH | 99.3%-100% | / | / |
| 2017[13] | (4 classes) from 44 patients by AAMI | 97.5% | / | / |
| 2016[2] | (5 classes) from 44 patients by AAMI | 92.7% | / | / |
| 2016[3] | from 44 patients in MIT-BIH | 98.9% | / | / |
| 2015[15] | ( Particular 5 classes) from 44 patients in MIT-BIH | 98.49% | / | / |
| 2012[16] | from 44 patients in MIT-BIH | 93.8% | / | / |

5. Conclusions

In this paper, we propose an ECG segmentation strategy of cutting ECG signals into segments for training and testing 1D CNN models of ECG signal classification. The experimental results demonstrate that the combination of this segmentation strategy and the 1D CNN model achieves an excellent classification accuracy, sensitivity, and F1-score. Meanwhile, the performance of our method is verified on all the recordings of all patients in the MIT-BIH. Specifically, as the heartbeat types are divided into 5 classes (i.e., normal beat, left bundle branch block beat, right bundle branch block beat, ventricular ectopic beat, and paced beat) in the MIT-BIH, the best classification accuracy, the AUC, the sensitivity, and the F1-score reach 99.24%, 0.9994, 0.99, and 0.99, respectively. As the heartbeat types are divided into 6 classes (i.e., normal beat, left bundle branch block beat, right bundle branch block beat, ventricular ectopic beat, paced beat, and other beat) in the MIT-BIH, the beat classification accuracy, the AUC, the sensitivity, and the F1-score reach 97.02%, 0.9966, 0.97, and 0.97, respectively. Meanwhile, as the heartbeat types are divided into 5 classes (i.e., normal beat, supraventricular ectopic beats, ventricular ectopic beats, fusion beats, and unclassifiable beats) by the AAMI recommendation, the beat classification accuracy, the sensitivity, and the F1-score reach 97.45%, 0.9925, and 0.97, respectively. The comprehensive evaluation has shown that our method achieves better performance than the state-of-the-art methods.

Applying recurrent neural networks to ECG classification is absolutely valuable because it has an inherent ability to process historical data like time series. Our future work will include exploration of the RNN model for ECG classification, and the automatic determination of the locations of R peaks.

6. Acknowledgment

This work was supported in part by the Key Lab of Computer Networks and Information Integration (Southeastern University), Ministry of Education, China (K93-9-2017-03), by the Department of Education Shaanxi Province, China (15JK1673), and by Shaanxi Provincial Natural Science Foundation of China (2016JM8034, 2018GY-135, 2018ZDXM-GY-091, 2020SF377). We gratefully acknowledge the support of NVIDIA Corporation with the donation of the Titan Xp GPU used for this research.